\documentclass[11pt,a4paper]{article}
\usepackage[utf8]{inputenc}
\usepackage{amsmath}
\usepackage{amsthm}
\usepackage{amssymb}
\usepackage{graphicx}

\makeatletter

\newcommand*\LyXZeroWidthSpace{\hspace{0pt}}

\newcommand{\lyxaddress}[1]{
	\par {\raggedright #1
	\vspace{1.4em}
	\noindent\par}
}
\theoremstyle{plain}
\newtheorem{thm}{\protect\theoremname}
\theoremstyle{plain}
\newtheorem{prop}[thm]{\protect\propositionname}

\@ifundefined{date}{}{\date{}}

\usepackage{enumitem}
\setlist[itemize]{label=-}

\usepackage{amsthm}
\usepackage{xcolor}

\makeatother

\providecommand{\propositionname}{Proposition}
\providecommand{\theoremname}{Theorem}

\begin{document}
\title{Nambu Nonequilibrium Thermodynamics and the Lyapunov Structure of Open Systems}
\author{So Katagiri\thanks{So.Katagiri@gmail.com} }
\maketitle

\lyxaddress{{\small\textit{Major in Complex Systems Science, Graduate School
of Science and Engineering, Ibaraki University, Nakanarusawa-cho,
Hitachi, 316-8511, Japan.}}}

\begin{abstract}
We investigate the thermodynamic meaning of the dissipation potential in Nambu nonequilibrium thermodynamics (NNET) using a simple open-piston model. The system is described by three variables: the gas volume, the piston momentum, and the thermodynamic entropy of the gas. The reversible dynamics is formulated in terms of two Nambu Hamiltonians, the extended energy and the thermodynamic entropy, while the irreversible dynamics is generated by a bath-relative Massieu-type dissipation potential through a positive-semidefinite gradient flow. This construction reproduces the physical processes of frictional heating and heat exchange with the reservoir while preserving their thermodynamic interpretation. We show that the time derivative of the dissipation potential coincides with the total entropy production rate of the system and the heat bath. Although the subsystem entropy may evolve nonmonotonically, the dissipation potential is a Lyapunov function under the conditions realized in this model. These results provide an explicit thermodynamic interpretation of the dissipation potential and its associated Lyapunov structure within NNET.
\end{abstract}

\section{Introduction}

\subsection{Entropy in Open Systems}

In Nambu nonequilibrium thermodynamics (NNET), a scalar dissipation potential is introduced to generate the irreversible part of the dynamics. However, its relation to the thermodynamic entropy of the system is not necessarily evident. In particular, in an open system exchanging heat and work with its surroundings, the thermodynamic entropy of the subsystem need not be monotonic, whereas a bath-relative thermodynamic potential may serve as a Lyapunov function.
\\

The thermodynamic entropy $S$ is a central concept in thermodynamics
and statistical mechanics, but its role is not unique. First, $S$
is a thermodynamic state variable, on the same footing as internal
energy and volume, and is connected to heat exchange through the first
law,

\begin{equation}
dU=TdS-PdV.
\end{equation}

Second, $S$ appears as the central quantity in the second law: in
an isolated system, it serves as an indicator of irreversibility,
or equivalently as a Lyapunov function characterizing the approach
to equilibrium. From the viewpoint of statistical mechanics, $S$
is also related to the logarithm of the number of microscopic states,
and hence to information, through Boltzmann’s relation
\begin{equation}
S=k_{B}\mathrm{ln}W.
\end{equation}

In equilibrium and isolated systems, these different roles are consistently
unified in a single quantity, the thermodynamic entropy $S$. In an
open system that exchanges heat and work with external reservoirs,
however, this unification is no longer automatic. Nevertheless, even
in relaxation processes of open systems, the picture that a certain
monotonic quantity increases as the system approaches equilibrium
remains valid under appropriate conditions. This monotonic quantity
is not necessarily the thermodynamic entropy $S$ itself, but may
be a different quantity constructed from it. The central issue addressed in this paper is to clarify how the dissipation potential introduced in NNET is related to thermodynamic entropy and to the Lyapunov structure of an open system.

\subsection{Challenges in Nonequilibrium Thermodynamics and the Description of
Temporal Order}

The thermodynamic interpretation of the dissipation potential in NNET is closely related to the broader problem of how reversible and irreversible dynamics are described within nonequilibrium thermodynamics. In particular, a consistent formulation should account for both the temporal evolution of a system, including oscillatory motion, and the thermodynamic constraints imposed by the first and second laws.
\\

Nonequilibrium thermodynamics began with individual empirical laws
established in the nineteenth century, and acquired a unified theoretical
framework in the twentieth century through Onsager’s theory of linear
response \cite{Onsager_1931}. By assuming detailed balance and linearity
near equilibrium, Onsager theory derived reciprocal relations among
transport coefficients and successfully described nonequilibrium phenomena
in terms of entropy production.
\\

Subsequently, Glansdorff and Prigogine formulated a general evolution
criterion for stationary states in the linear regime, thereby advancing
the understanding of dissipative structures and self-organization
in open systems \cite{Glansdorff_1964}.
\\

However, these frameworks alone do not make it
straightforward to describe concrete time evolution
in nonlinear far-from-equilibrium states, especially
forms of temporal order such as periodic motion and
rotational flows, within a thermodynamic framework.

Variational principles for the nonlinear regime have
also been proposed, including the maximum entropy
production principle of the Ziegler type
\cite{Ziegler1963}.
The combination of conservative and dissipative
dynamics has been extensively studied in generalized
bracket formulations, including the Beris--Edwards
formalism and GENERIC
\cite{Edwards_1991,Grmela_1997,Ottinger_1997}.
Related Nambu-metriplectic formulations have also
been developed for dissipative fluid systems,
including Rayleigh--B\'enard convection
\cite{Bihlo_2008}.

In NNET, we employ the Nambu bracket to describe
the circulating dynamics directly in terms of
multiple physically identified invariants.
Although such dynamics can also be represented
in Poisson form, a Poisson reformulation is not
required for the present construction.
The Nambu formulation makes the roles of the
multiple invariants explicit and symmetric.
The present work focuses on the thermodynamic
meaning of the dissipation potential introduced
in this Nambu-based framework.
\\

Oscillations and periodic motion observed in nonequilibrium systems
are naturally viewed as the coexistence of conservative rotational
flows and dissipative relaxation flows. Recent developments in stochastic
thermodynamics and active matter physics have drawn attention to probability
currents generated by nonconservative forces and to the geometric
structure of entropy production. In stochastic thermodynamics, frameworks
have been developed for decomposing entropy production geometrically,
and for structurally decomposing probability currents into gradient
and nongradient components. In particular, Dechant, Sasa, and Ito
\cite{Dechant_2022} and Yoshimura et al. \cite{Yoshimura_2023} showed
that probability currents in nonequilibrium systems can be decomposed
into a gradient-flow component associated with relaxation and a nongradient
component associated with steady cycle currents. However, at the level
of macroscopic phenomenological thermodynamics, frameworks that directly
treat the geometric structure of the reversible part, namely rotational
flows, remain limited. This observation motivates a macroscopic formulation
in which reversible rotational dynamics and dissipative gradient relaxation
are separated explicitly.
In such a formulation, an important question is whether the scalar potential generating the dissipative flow can be assigned a definite thermodynamic meaning.

\subsection{Nambu Nonequilibrium Thermodynamics and the Aim of This Paper}

The present work is intended as a continuation of the macroscopic
tradition of nonequilibrium thermodynamics initiated by Onsager and
further developed by Glansdorff and Prigogine. Its purpose is not
to reduce irreversible thermodynamics to microscopic statistical mechanics,
but to extend the macroscopic description so that reversible temporal
order and irreversible relaxation can be represented within a common
geometric framework.
\\

The central aim of this paper is therefore to clarify the thermodynamic meaning of the dissipation potential introduced in Nambu nonequilibrium thermodynamics and to examine how its Lyapunov structure emerges in a concrete open thermodynamic system.
\\

Motivated by this perspective, we have proposed Nambu nonequilibrium
thermodynamics (NNET) as a macroscopic framework for treating reversible
and irreversible parts in a covariant and unified manner \cite{Katagiri_2022_Fluctuating,Katagiri_2025_Nambu}.
In NNET, the reversible part is described by the Nambu bracket, namely
a generalized Hamiltonian dynamics, whereas the irreversible part
is represented as a gradient flow generated by a scalar function $S_{NB}$.
The mathematical background of this theory is provided by the Helmholtz
decomposition and Darboux’s theorem \cite{Katagiri_2026}. This structure
provides a possible route to describing dissipative systems with periodic
dynamics, such as chemical reaction systems and neuronal models \cite{Katagiri_2025_Applications}.
\\

Previous studies introduced NNET as a geometric framework for decomposing
nonlinear nonequilibrium dynamics into Nambu-type reversible flow
and dissipative gradient flow. The present paper focuses on a more
elementary but fundamental issue: the thermodynamic meaning of the
dissipation potential itself in an open macroscopic system.
\\

Throughout this paper, the term ``adiabatic reversible piston'' refers to
an ideal reference dynamics in which all irreversible relaxation channels
are suppressed. This is not meant to describe a real adiabatic piston with
friction, viscosity, acoustic damping, or other internal relaxation mechanisms.
In a real adiabatic piston, entropy may be produced even though no heat is
exchanged with the exterior. Such irreversible effects are introduced
separately in the dissipative sector.
\\

The main contributions are as follows.
\\

First, we formulate the reversible piston dynamics as a Nambu rotational
flow generated by the extended energy and the subsystem entropy. This
shows that the mechanical oscillation is not part of the dissipative
sector, but belongs to the reversible geometric structure.
\\

Second, we construct a dissipative gradient-flow sector generated
by $S_{NB}=S-H_{1}/T_{b}$, where $H_{1}$ is the extended energy
including the pressure-reservoir work term. This identifies $S_{\mathrm{NB}}$ as a bath-relative Massieu-type potential, distinct from the subsystem thermodynamic entropy, and shows that the resulting gradient flow reproduces frictional heating and heat exchange consistently with the thermodynamic balance laws.
\\

Third, we prove a conditional Lyapunov criterion: $S_{NB}$ is monotonic
when it is preserved by the reversible sector and generates a positive-semidefinite
irreversible gradient flow. Numerical simulations then show the coexistence
of damped piston oscillations, nonmonotonic subsystem entropy, and
monotonic $S_{NB}$. In this model, the increase of $S_{\mathrm{NB}}$ coincides with the total entropy increase of the subsystem and the ideal heat bath, providing an explicit thermodynamic interpretation of the Lyapunov structure.
\\

The present formulation should be viewed as complementary to Onsager
theory and GENERIC. While Onsager theory emphasizes near-equilibrium
linear response and GENERIC ensures thermodynamic consistency through
degeneracy conditions, the present model emphasizes a directly visible
separation between reversible rotational motion and dissipative gradient
relaxation\cite{Katagiri_2026}.
\\

The remainder of this paper is organized as follows. In Sec. 2, we
summarize the basic structure of Nambu nonequilibrium thermodynamics.
In Sec. 3, we show that an adiabatic reversible piston system coupled
to a pressure bath can be naturally formulated in the Nambu form.
In Sec. 4, we introduce a heat bath and friction, construct the irreversible
part, and discuss the relation between the dissipation potential and
entropy production. In Sec. 5, we present numerical results. Finally,
in Sec. 6, we discuss possible extensions to chemical reaction systems
and the broader significance of the present framework.

\footnote{It should be noted that, in port-thermodynamic systems based on contact
geometry and symplectization, a gas-piston-damper system has been
described as homogeneous Hamiltonian/contact dynamics on an extended
space $(q,S,E,p,p_{S},p_{E})$\cite{Van_der_Schaft_2018}. In contrast,
the present work describes the reversible part by the Nambu bracket
and the irreversible part by the gradient flow of $S_{NB}$\LyXZeroWidthSpace{}
directly on the original macroscopic state variables $(V,p,S)$, thereby
allowing a direct comparison between $S_{NB}$ and the thermodynamic
entropy $S$.}

\section{Basic Structure of Nambu Nonequilibrium Thermodynamics}

Let the state variables of the system be denoted by 
\begin{equation}
x=(x^{1},\dots,x^{N}).
\end{equation}
In Nambu nonequilibrium thermodynamics (NNET), the time evolution
is decomposed into reversible and irreversible parts\cite{Katagiri_2025_Nambu}:
\begin{equation}
\dot{x}^{i}=\partial_{t}^{(\mathrm{H})}x^{i}+\partial_{t}^{(\mathrm{S})}x^{i}.
\end{equation}

Here, $\partial_{t}^{(H)}x^{i}$ and $\partial_{t}^{(S)}x^{i}$ denote
the reversible and irreversible parts of the time evolution, respectively.

\subsection{Nambu Bracket and the Reversible Part}

The reversible part is formulated in terms of $N-1$ functions 
\begin{equation}
H_{1},\dots,H_{N-1}
\end{equation}
which are invariants of the reversible sector, and the Nambu bracket\cite{Katagiri_2025_Nambu}.
The Nambu bracket, introduced by Yoichiro Nambu in 1973\cite{Nambu_1973}
as a generalization of Hamiltonian mechanics, is the completely antisymmetric
Jacobian defined by
\begin{equation}
\{A_{1},\dots,A_{N}\}\equiv\epsilon^{i_{1}\cdots i_{N}}\frac{\partial A_{1}}{\partial x^{i_{1}}}\cdots\frac{\partial A_{N}}{\partial x^{i_{N}}}.
\end{equation}
Using this bracket, the reversible part of the dynamics can be written
as 
\begin{equation}
\partial_{t}^{(\mathrm{H})}x^{i}=\{x^{i},H_{1},\dots,H_{N-1}\}
\end{equation}
Because of the complete antisymmetry of the Nambu bracket, one immediately
obtains
\begin{equation}
\partial_{t}^{(H)}H_{m}=0.
\end{equation}
Thus, $H_{m}$ are conserved by the reversible dynamics. In the full
dynamics, however, their conservation depends on the structure of
the irreversible sector introduced below; see Ref.\cite{Katagiri_2025_Nambu}.
In NNET, the Nambu sector is characterized by
its geometric structure and conservation
properties, and is not assumed to be
time-reversal invariant in general.
As discussed by Pavelka et al. in the context
of GENERIC, the identification of a Hamiltonian
sector with time-reversal reversible dynamics
requires additional conditions
\cite{Pavelka_2014}.
In the particular piston model considered here,
the Nambu sector is time-reversal invariant and
represents the reversible adiabatic reference
dynamics.

\begin{figure}[h]
\centering
\includegraphics[width=0.5\textwidth,totalheight=0.5\textheight,keepaspectratio]{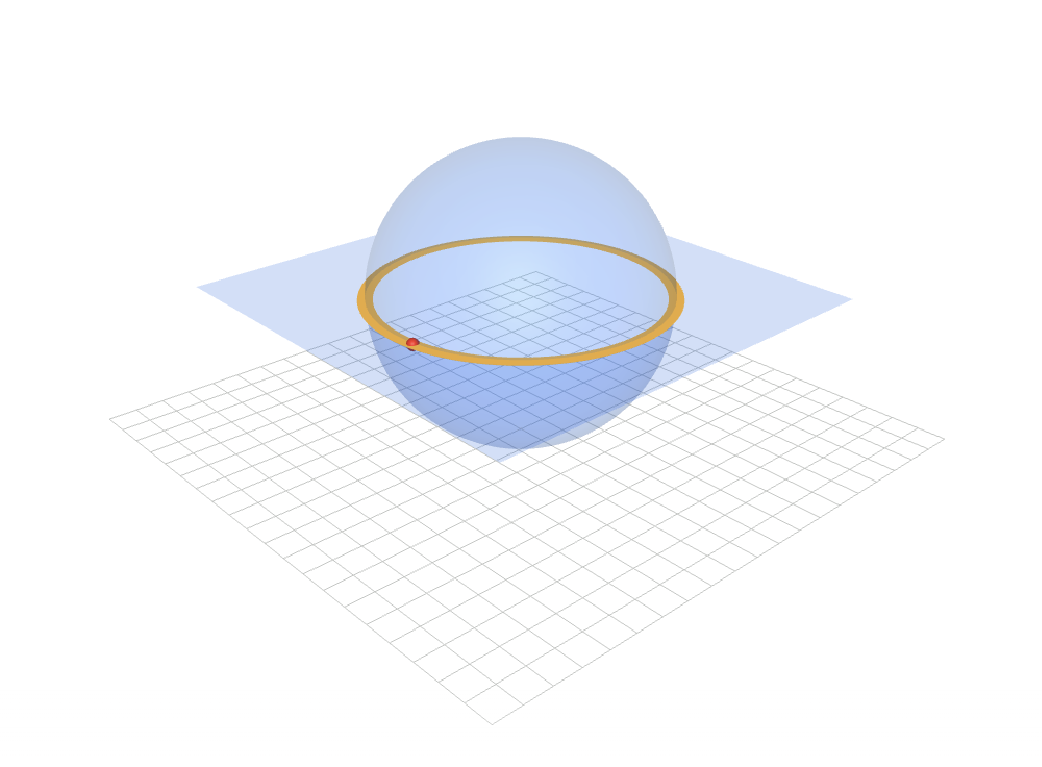}\caption{Schematic illustration of the reversible Nambu flow for $N=3$. At
regular points, the intersection of the two level surfaces is locally
one-dimensional. In the present schematic example, the intersection
is drawn as a closed curve.}\label{fig:Schematic-illustration-of}
\end{figure}

Furthermore, in a regular region where
\begin{equation}
\nabla H_{1},\dots,\nabla H_{N-1}
\end{equation}
are linearly independent, the common intersection of the $N-1$ level
surfaces is locally a one-dimensional curve, as schematically illustrated
in Fig.~\ref{fig:Schematic-illustration-of}. The reversible vector
field generated by the Nambu bracket is tangent to this curve. In
this sense, it represents a circulating, or rotational, flow along
the intersection of the conserved level surfaces. In particular, when
this intersection forms a closed curve, the motion can be interpreted
as periodic motion, in the same manner as in a harmonic oscillator.

\subsection{Dissipation Potential and the Irreversible Part}

The irreversible part of NNET is generally written in the following
gradient-flow form using a transport coefficient matrix $L^{ij}$\cite{Katagiri_2025_Nambu}:
\begin{equation}
\partial_{t}^{(\mathrm{S})}x^{i}=L^{ij}\frac{\partial S_{NB}}{\partial x^{j}}.
\end{equation}
Here, $L^{ij}$ can be decomposed into symmetric and antisymmetric
parts. The structure in which irreversible dynamics is described as
a gradient flow generated by a thermodynamic potential has recently
been reinterpreted from the viewpoint of information geometry. In
particular, attempts have been made to understand Onsager’s nonequilibrium
thermodynamics as a natural gradient flow on the space of probability
distributions\cite{Wada_2025}. As another geometric direction, formulations
have also been proposed in which thermodynamic forces are regarded
as gauge fields, and Onsager-type constitutive relations are derived
as gauge-fixing conditions. In such gauge-fixing theories, nonlinear
constitutive relations and transverse components appear by extending
thermodynamic forces from pure gauge fields to physical gauge fields
\cite{Katagiri_2018,Aibara_2019}. In the present paper, however,
we do not introduce such transverse components as thermodynamic gauge
fields. Instead, we explicitly separate the reversible nongradient
component as a dynamical sector generated by the Nambu bracket.
\\

In this paper, in order to describe the irreversible part as a gradient
flow of $S_{NB}$, we use the symmetric positive-semidefinite part
of the transport coefficient matrix $\ensuremath{L^{ij}}$. The reversible
rotational flow is not treated as an antisymmetric component of the
irreversible part, but is described separately by the Nambu bracket.
This separation clarifies the distinct roles of the reversible and
irreversible sectors\footnote{In Heimburg's formulation\cite{Heimburg_2017}, isentropic oscillations
are generated by the antisymmetric part of an Onsager-type linear
phenomenological coefficient. In the terminology of the present paper,
this corresponds to the near-equilibrium linearization of the reversible
sector. Here, the reversible part is separated from the outset as
a geometric rotational flow generated by the Nambu bracket, rather
than being included as an antisymmetric component of the irreversible
transport coefficient. In the present formulation, the reversible motion is already represented by the Nambu bracket. Including the same reversible contribution again through antisymmetric cross terms would therefore result in double counting.}.
\\

$S_{NB}$ is a scalar function that generates the irreversible part
of the dynamics, and is referred to here as the dissipation potential.
In general, it does not coincide with the thermodynamic entropy $S$.
A pure gradient flow alone cannot generate periodic motion. In NNET,
however, temporal order in far-from-equilibrium systems can be naturally
represented through the coupling of the irreversible gradient flow
to the reversible part, namely the rotational flow generated by the
Nambu bracket.
\\

The central issue addressed in the present paper is the thermodynamic meaning of this dissipation potential and its relation to the Lyapunov structure of an open system. The
quantity $S_{NB}$ is not the thermodynamic entropy $S$ itself that
appears in the second law, but the generating potential used to represent
the irreversible vector field as a gradient flow. Therefore, the monotonicity
of $S_{NB}$ is not an automatic or universal consequence of its definition.
For $S_{NB}$ to function as a Lyapunov quantity, the reversible part
must preserve $S_{NB}$, and the irreversible part must close as a
gradient flow of $S_{NB}$ with a positive-semidefinite transport
coefficient. The piston model studied in this paper provides a minimal
example in which these conditions are realized explicitly.
In Morrison's metriplectic formulation,
the generalized entropy is constructed as
a function of Casimir invariants of the
Poisson bracket \cite{Morrison_1986}.
In NNET, by contrast, the dissipation
potential is not generally required to be
invariant under the Nambu flow.
Its preservation by the Nambu sector is
imposed here only as a sufficient condition
for the Lyapunov property examined in the
present piston model.

\section{Nambu Dynamics of an Adiabatic Reversible Piston System}

In this section, we consider the adiabatic reversible process of an
ideal piston system without friction.
The purpose of this section is to isolate the reversible reference sector
that will later be combined with irreversible relaxation. It is not intended
to describe a real adiabatic piston by itself. A real adiabatic piston may
produce entropy through friction, viscosity, acoustic damping, or internal
thermal relaxation, even when the walls are adiabatic. Those effects are
excluded in the present section and are introduced in Sec. 4 as the
irreversible sector.
 We show that this process is
naturally described by the Nambu bracket and that it contains an intrinsic
oscillatory mode.

\subsection{State Variables and Thermodynamic Quantities}

\begin{figure}
\centering
\includegraphics[width=1\textwidth,totalheight=0.5\textheight,keepaspectratio]{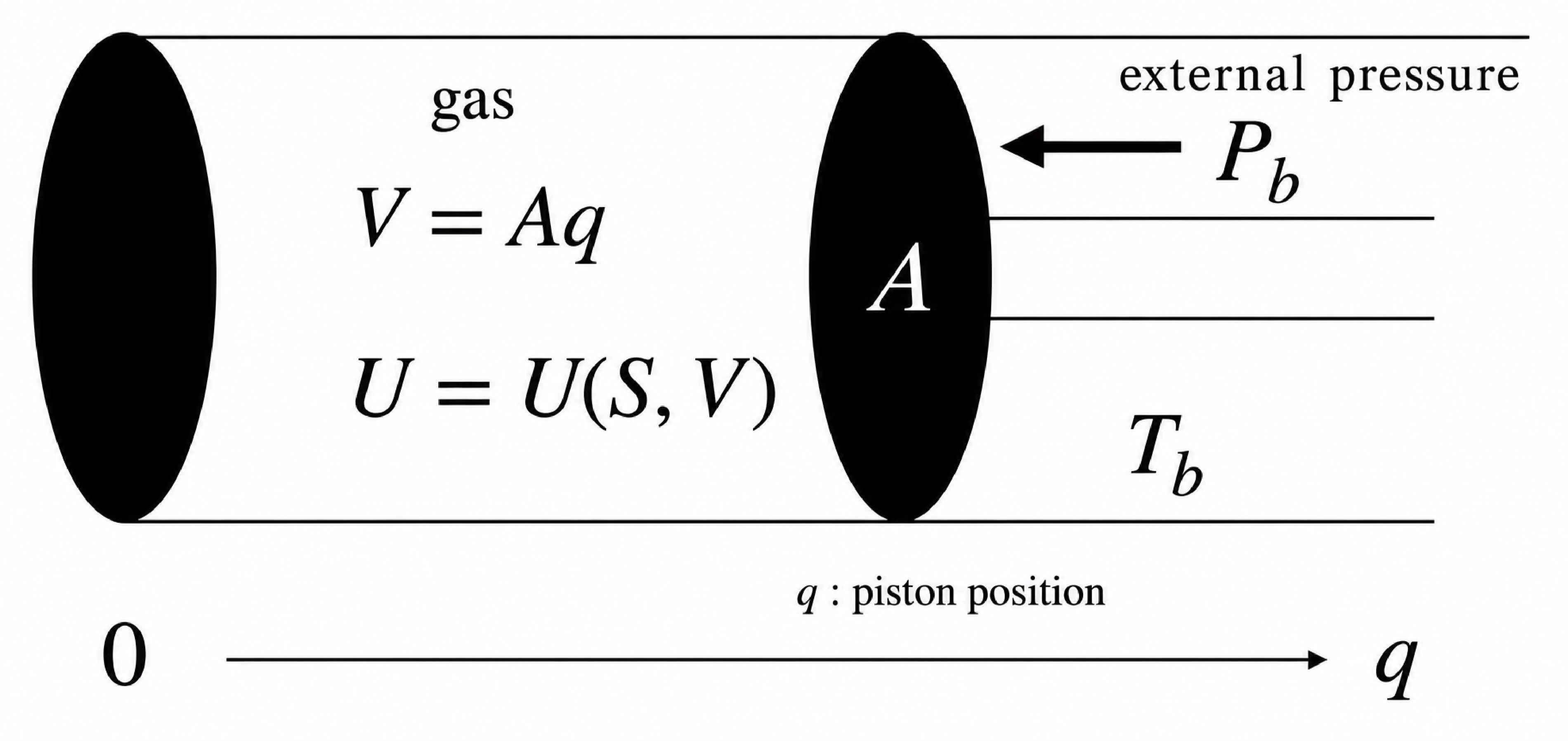}

\caption{Piston system considered in this paper. The symbol \(T_b\)
indicates the heat bath introduced only in Sec. 4. In Sec. 3, only the
ideal adiabatic reversible reference sector is considered, and all
irreversible relaxation channels, including heat exchange and friction,
are suppressed.}
\label{fig:piston-system}

\end{figure}

Let $A$ be the cross-sectional area of the piston and $q$ its position,
so that the gas volume is given by

\begin{equation}
V=Aq.
\end{equation}

We define the physical momentum of the piston by
\begin{equation}
p=m\dot{q},
\end{equation}
which gives
\begin{equation}
\dot{V}=\frac{A}{m}p.
\end{equation}

Let $P_{b}$ denote the external pressure, $S$ the thermodynamic
entropy of the gas, and $U(S,V)$ its internal energy. The thermodynamic
relations are then 
\begin{equation}
T=\frac{\partial U}{\partial S},\qquad P=-\frac{\partial U}{\partial V}.
\end{equation}
In the ideal adiabatic reversible sector considered in this section, there
is no entropy production and hence
\begin{equation}
\dot{S}=0 .
\end{equation}
The corresponding frictionless piston equation is
\begin{equation}
\dot{p}=A(P-P_{b}).
\end{equation}
This equation is not meant as the most general effective equation for a
real adiabatic piston. If irreversible internal relaxation or friction is
retained, a dissipative term such as \(-\lambda p\) must be added, and the
lost kinetic energy contributes to entropy production. Such terms are
introduced in Sec. 4.

It is therefore natural to choose the three state variables of the
system\footnote{The present model should be understood as a lumped macroscopic description
in which the gas is assumed to remain sufficiently close to local
equilibrium and is represented only by the global variables $V$,
$p$, and $S$. Spatially inhomogeneous hydrodynamic modes, such as
vortical flows, viscous boundary layers, acoustic transients, and
temperature gradients inside the gas, are not included explicitly.
This approximation is appropriate when these internal relaxation modes
decay on time scales shorter than, or are otherwise coarse-grained
over, the piston relaxation time considered here. If such modes become
comparable to the piston time scale, the state space must be enlarged
to include hydrodynamic fields, and the present minimal model should
be regarded as the reduced slow-variable description.} as

\begin{equation}
x=(V,p,S).
\end{equation}
Here, because we use the volume $V=Aq$ and the physical momentum
$\ensuremath{p=m\dot{q}}$ as state variables, the pair $(V,p)$ is
not a canonical pair. Indeed, for the canonical pair$(q,p)$, the
Poisson bracket satisfies

\begin{equation}
\ensuremath{\{V,p\}=A}.
\end{equation}
Therefore, in the coordinates $\ensuremath{(V,p,S)}$ , we use the
normalized Nambu bracket
\begin{equation}
\{F,G,K\}_{A}\equiv A\{F,G,K\}\equiv A\frac{\partial(F,G,K)}{\partial(V,p,S)}.
\end{equation}

\subsection{Invariants of the Reversible Sector and Nambu Formulation}

In the ideal reversible reference sector, the Nambu dynamics has two
invariants. The first is the extended energy \(H_1\), which includes the
work exchange with the pressure reservoir. The second is
\(H_2=S\), the thermodynamic entropy of the gas, which is invariant only
within this reversible sector.

This statement should not be understood as a claim that the entropy of a
real adiabatic piston with friction or internal relaxation is conserved.
For such an irreversible adiabatic piston, \(H_1\) may remain conserved
while \(S\) increases. The role of the present section is only to identify
the reversible Nambu part of the dynamics.

The two invariants of the reversible sector are therefore written as
\begin{equation}
H_{1}=\frac{p^{2}}{2m}+U(S,V)+P_{b}V,\qquad H_{2}=S.
\end{equation}

Here and below, the superscript \((H)\) indicates that the time derivative
is evaluated only along the reversible Nambu sector. It does not imply that
the same conservation laws hold after the irreversible sector is added.

Indeed, using the reversible equations of motion, one obtains
\begin{equation}
\dot{H}_{1}^{(\mathrm{H})}=\frac{p}{m}\dot{p}+\partial_{V}U\,\dot{V}+P_{b}\dot{V}=0.
\end{equation}

Taking these quantities as the Hamiltonians of the Nambu dynamics,
the time evolution generated by the normalized Nambu bracket becomes
\begin{equation}
\partial_{t}^{(\mathrm{H})}V=\{V,H_{1},H_{2}\}_{A}=A\frac{\partial(V,H_{1},S)}{\partial(V,p,S)}=\frac{Ap}{m},
\end{equation}
\begin{equation}
\partial_{t}^{(\mathrm{H})}p=\{p,H_{1},H_{2}\}_{A}=A\frac{\partial(p,H_{1},S)}{\partial(V,p,S)}=-A(\partial_{V}U+P_{b})=A(P-P_{b}),
\end{equation}
and

\begin{equation}
\partial_{t}^{(\mathrm{H})}S=\{S,\ H_{1},H_{2}\}_{A}=0.
\end{equation}
Thus, the Nambu formulation exactly reproduces the expected reversible
dynamics of the adiabatic piston system.

\subsection{Linearization Around Equilibrium and the Oscillatory Mode}

To show that the oscillatory motion is generated by the reversible
part, we linearize the dynamics around an equilibrium point
\begin{equation}
(V_{*},p_{*}=0,S_{*}).
\end{equation}
The equilibrium condition is 
\begin{equation}
P(S_{*},V_{*})=P_{b}.
\end{equation}
Writing
\begin{equation}
V=V_{*}+\delta V,p=\delta p
\end{equation}
we obtain
\begin{equation}
\frac{d\delta V}{dt}=\frac{A\delta p}{m},\qquad\frac{d\delta p}{dt}=-A\left.\frac{\partial^{2}U}{\partial V^{2}}\right|_{*}\delta V.
\end{equation}
These equations give the harmonic-oscillator equation 
\begin{equation}
\dfrac{d^{2}\delta V}{dt^{2}}+\omega_{*}^{2}\,\delta V=0,\qquad\omega_{*}^{2}=\frac{A^{2}}{m}\left.\frac{\partial^{2}U}{\partial V^{2}}\right|_{*}.
\end{equation}
The eigenvalues are therefore
\begin{equation}
\mu_{\pm}=\pm i\omega_{*}.
\end{equation}

Thus, the oscillatory mode near equilibrium, namely the temporal order
of the piston motion, originates purely from the reversible part generated
by the Nambu bracket.

\section{Extension to an Open System: Heat Bath and Irreversible Dynamics}

In real systems, dissipation and heat exchange with the environment
are present. In this section, we introduce a heat bath at temperature
$T_{b}$ and piston friction, and clarify the role of the irreversible
part.

It is useful to note that the adiabatic but irreversible piston is recovered
as the special case \(\kappa=0\) and \(\lambda>0\). In that case the system
does not exchange heat with an external heat bath, but friction converts
piston kinetic energy into internal heat. Consequently, the extended energy
$H_1$ is conserved, whereas the entropy satisfies
\begin{equation}
\dot{S} = \frac{\lambda p^2}{mT} \ge 0 .
\end{equation}
Thus the present formulation explicitly includes the standard irreversible
adiabatic piston; it is not assumed that friction appears only through
coupling to a heat bath.

\subsection{Phenomenological Model and Entropy Production}

We consider an explicit model in which the kinetic energy dissipated
by friction is converted into heat within the system. The irreversible
part is introduced as 
\begin{equation}
\partial_{t}^{(\mathrm{S})}V=0,\qquad\partial_{t}^{(\mathrm{S})}p=-\lambda p,\qquad T\,\partial_{t}^{(\mathrm{S})}S=\kappa(T_{b}-T)+\frac{\lambda}{m}p^{2}.
\end{equation}
Here, $\lambda$ is the coefficient characterizing piston friction,
and determines the decay rate of the kinetic energy. The second term
in $T\partial_{t}^{(S)}S$ represents the conversion of the kinetic
energy lost by friction into internal heat\footnote{Strictly speaking, the frictional heat may be distributed between
the gas, the piston, and the external reservoir, depending on the
microscopic realization of the contact. In the present minimal model,
the thermal degrees of freedom of the piston are not included explicitly,
and we assume that the kinetic energy dissipated by friction is effectively
recovered as heat in the gas subsystem. If part of this heat is instead
transferred directly to the heat bath, the bath temperature is still
kept fixed in the ideal-reservoir limit; such a redistribution would
correspond to a different coarse-grained heat partition and does not
affect the geometric point addressed here. In particular, the existence of the friction channel does not rely on the
presence of the heat bath; the heat bath only provides an additional thermal
relaxation channel.}. Indeed, the kinetic energy lost by friction is

\begin{equation}
\frac{\partial}{\partial t}\left(\frac{p^{2}}{2m}\right)=\frac{p\dot{p}}{m}=-\frac{\lambda}{m}p^{2}.
\end{equation}
The coefficient $\kappa>0$ is a phenomenological parameter characterizing
heat exchange with the heat bath, corresponding to Newton's law of
cooling.

With the irreversible terms introduced above, the energy change of
the system follows from the total time evolution 
\begin{equation}
\dot{O}=\partial_{t}^{(\mathrm{H})}O+\partial_{t}^{(\mathrm{S})}O.
\end{equation}
Since $\partial_{t}^{(H)}H_{1}=0,$ we obtain
\begin{equation}
\dot{H_{1}}=\partial_{t}^{(H)}H_{1}+\partial_{t}^{(S)}H_{1}=\partial_{t}^{(S)}H_{1}=-\lambda\frac{p^{2}}{m}+T\partial_{t}^{(S)}S=\kappa(T_{b}-T).
\end{equation}
If the energy change rate of the heat bath is defined as
\begin{equation}
\dot{E}_{b}=-\kappa(T_{b}-T),
\end{equation}
then
\begin{equation}
\dot{H}_{1}+\dot{E}_{b}=0.
\end{equation}
Thus, the first law, namely energy conservation for the system plus
bath, is satisfied.

\LyXZeroWidthSpace{}

We next discuss the second law. In this paper, the pressure reservoir
is treated as an ideal work reservoir: it exchanges work with the
system but carries no thermodynamic entropy, so that $\dot{S}_{P}=0$.
Under this idealization, the total entropy production, given by the
entropy change of the system plus that of the heat bath, becomes 
\begin{equation}
\dot{S}_{\mathrm{tot}}=\dot{S}+\frac{\dot{E}_{b}}{T_{b}}=\kappa\frac{(T_{b}-T)^{2}}{TT_{b}}+\frac{\lambda}{mT}p^{2}\ge0.
\end{equation}
This confirms consistency with the second law of thermodynamics. It
should be noted, however, that the thermodynamic entropy $S$ refers
only to the open subsystem, and therefore its time derivative is allowed
to become negative.

\subsection{Gradient-Flow Representation by the Dissipation Potential $S_{NB}$ }

To organize the phenomenological model introduced above within the
framework of NNET, we define the dissipation potential $S_{NB}$ ,
which generates the irreversible part, as 
\begin{equation}
S_{NB}=S-\frac{H_{1}}{T_{b}}=S-\frac{1}{T_{b}}\left(U(S,V)+\frac{p^{2}}{2m}+P_{b}V\right).
\end{equation}
The gradient of $S_{\mathrm{NB}}$ is then
\begin{equation}
\frac{\partial S_{\mathrm{NB}}}{\partial V}=\frac{P-P_{b}}{T_{b}},\qquad\frac{\partial S_{\mathrm{NB}}}{\partial p}=-\frac{p}{mT_{b}},\qquad\frac{\partial S_{\mathrm{NB}}}{\partial S}=1-\frac{T}{T_{b}}=\frac{T_{b}-T}{T_{b}}.
\end{equation}
The form of $S_{NB}$ adopted here is not an arbitrary assumption.
Since the system is in contact with a heat bath at temperature $T_{b}$
and a work reservoir at pressure $P_{b}$, it is natural, under the
fixed external conditions $(T_{b},P_{b})$, to construct a free-energy-like
potential from the extended energy
\begin{equation}
H_{1}=U(S,V)+\frac{p^{2}}{2m}+P_{b}V,
\end{equation}
which includes the internal energy, the kinetic energy of the piston,
and the work term associated with the pressure reservoir. Thus,
\begin{equation}
S_{NB}=S-\frac{H_{1}}{T_{b}}
\end{equation}
can be interpreted as the negative of the generalized free energy
$H_{1}-T_{b}S$, divided by $T_{b}$\LyXZeroWidthSpace . It is therefore
an effective potential with the dimension of entropy. This is precisely
the point: $S_{NB}$ is not the thermodynamic entropy itself, but
a Massieu/free-energy-like potential adapted to the imposed bath variables\footnote{The temperature appearing in this potential is the reservoir temperature
$T_{b}$, not the instantaneous system temperature $T$. Thus $S_{NB}$
should be understood as a bath-relative Massieu potential under the
imposed external condition $T_{b}=\mathrm{const\ensuremath{}}$.,
rather than as an approximation obtained by replacing $T$ with $T_{b}$.
The deviation of the system temperature from the bath temperature
appears instead as the thermodynamic force $\partial S_{NB}/\partial S=1-T/T_{b}$.}. 

The bath-relative Massieu-type potential used
here is closely related to the conventional
thermodynamic availability or exergy employed
in the stability analysis of open systems,
as discussed, for example, by Hoang et al.
\cite{Hoang_2013}.
We do not claim novelty for this thermodynamic
potential itself.
Rather, the purpose of the present construction
is to establish its explicit identification
with the dissipation potential of NNET and
to demonstrate how the associated gradient
flow reproduces the physical irreversible
processes of the piston system.

To represent both the frictional decay of the momentum and the recovery
of the lost kinetic energy as internal heat within the gradient flow
of $S_{NB}$, we adopt the following positive-semidefinite transport
coefficient matrix.

The transport coefficient matrix can be decomposed into two physically
distinct dissipative channels:

\begin{equation}
L^ {}=L_{\mathrm{fric}}^ {}+L_{heat}^ {}.
\end{equation}

\subsubsection*{friction channel}

\begin{equation}
L_{\mathrm{fric}}^ {}=\lambda|v\rangle\langle v|
\end{equation}

\begin{equation}
|v\rangle=\sqrt{mT}|p\rangle-\frac{p}{\sqrt{mT}}|S\rangle
\end{equation}

This rank-one contribution describes the conversion of kinetic energy
into internal heat.

\subsubsection*{heat-bath channel
\begin{equation}
L_{\mathrm{heat}}=\kappa\frac{T_{b}}{T}|S\rangle\langle S|.
\end{equation}
}

This contribution represents thermal relaxation toward the bath temperature.

\subsubsection*{positivity}

Since both contributions are positive semidefinite for

\begin{equation}
T>0,\ \lambda\geq0,\ \kappa\geq0,
\end{equation}
their sum is positive semidefinite:

\begin{equation}
\langle x|\left(L_{\mathrm{fric}}+L_{\mathrm{heat}}\right)|x\rangle=\lambda\langle x|v\rangle\langle v|x\rangle+\kappa\frac{T_{b}}{T}\langle x|S\rangle\langle S|x\rangle\geq0.
\end{equation}

Then one obtains
\begin{equation}
\partial_{t}^{(S)}V=0,\qquad\partial_{t}^{(S)}p=-\lambda p,\qquad\partial_{t}^{(S)}S=\kappa\frac{T_{b}-T}{T}+\frac{\lambda p^{2}}{mT}.\label{eq:V,p,SHeatTimeVari}
\end{equation}
Thus, the phenomenological irreversible dynamics introduced in the
previous subsection is reproduced.

\LyXZeroWidthSpace{}

It is important to note that the reversible part in the present formulation
has already been specified by the Nambu bracket,
\begin{equation}
\partial_{t}^{(H)}x^{i}=\{x^{i},H_{1},H_{2}\}_{A}.
\end{equation}

Therefore, it is not necessary to introduce Onsager-{}-Casimir-type
antisymmetric cross terms into the irreversible part. In the irreversible
sector of the present model, we use only the symmetric positive-semidefinite
part of the transport coefficient matrix in order to ensure the monotonicity
of $S_{\mathrm{NB}}$.

The time evolution of $S_{NB}$ is determined solely by the irreversible
part, since the reversible contribution satisfies

\begin{equation}
\partial_{t}^{(H)}S_{NB}=0.
\end{equation}
Therefore,
\begin{equation}
\dot{S}_{\mathrm{NB}}=\partial_{t}^{(S)}S_{\mathrm{NB}}=\frac{\lambda p^{2}}{mT}+\kappa\frac{(T-T_{b})^{2}}{TT_{b}}\ge0.\label{eq:SNBTimeVari}
\end{equation}

The equality $\dot{S}_{NB}=0$ is attained only when both dissipative
channels are inactive. For $\lambda>0$, $\kappa>0$, and $T>0$,
Eq. \eqref{eq:SNBTimeVari} implies $p=0$ and $T=T_{b}$, namely
the piston has no kinetic motion and the gas is in thermal equilibrium
with the heat bath. At a fixed point of the full dynamics, the mechanical
equation further requires $P=P_{b}$. Thus the equilibrium state is
characterized by
\begin{equation}
p=0,\qquad T=T_{b},\qquad P=P_{b}.
\end{equation}

In this state, the system entropy also becomes stationary, since Eq.
\eqref{eq:V,p,SHeatTimeVari} gives $\dot{S}=0$.

Thus, whereas the thermodynamic entropy $S$ of the system may oscillate,
the dissipation potential $S_{NB}$ functions as a monotonically increasing
Lyapunov function in the explicit dissipative model adopted in this
section.

This monotonicity is analogous to a corresponding geometric structure
in stochastic thermodynamics. In Fokker-{}-Planck dynamics, the Kullback-{}-Leibler
divergence between probability distributions is known to decrease
monotonically along a natural gradient flow, and this property has
been rigorously established within geometric thermodynamics based
on information geometry and optimal transport theory\cite{Ito_2023}.
The monotonic increase of $S_{NB}$ in the present model may be regarded
as a macroscopic counterpart of this structure.

\LyXZeroWidthSpace{}

However, this monotonic increase does not follow unconditionally from
the definition of $S_{NB}$.

\LyXZeroWidthSpace{}

In the present model, the monotonicity above holds because $S_{NB}$
is constructed from the conserved quantity $H_{1}$ and the thermodynamic
entropy $S$, the reversible part leaves $S_{NB}$ unchanged, and
the irreversible part closes as a gradient flow of $S_{NB}$ with
a positive-semidefinite transport coefficient. In more general strongly
nonlinear far-from-equilibrium systems, for example when the reversible
part does not preserve $S_{NB}$, or when the irreversible part does
not close as a positive-semidefinite gradient flow generated by a
single $S_{NB}$, $S_{NB}$ itself may exhibit nonmonotonic oscillations.
Therefore, the result obtained here is not a claim of universal monotonicity
of $S_{NB}$, but rather demonstrates, in the explicit piston model,
sufficient conditions under which $S_{NB}$ functions as a Lyapunov
quantity.

The following proposition does not state that $S_{\mathrm{NB}}$ is
always a Lyapunov function. Rather, it states that the monotonicity
of $S_{\mathrm{NB}}$ is guaranteed under specific geometric conditions,
such as those realized in the present piston model.
The distinction between invariants of the reversible sector and monotonic
quantities of the full irreversible dynamics is essential. The entropy
\(S\) is an invariant only of the reversible sector, whereas the monotonic
quantity in the open dissipative model is \(S_{\rm NB}\), under the
conditions stated below.

\begin{prop}
Conditional Lyapunov property of $S_{\mathrm{NB}}$.

\;

Suppose that the irreversible part can be written as
\begin{equation}
\partial_{t}^{(S)}x^{i}=L^{ij}\frac{\partial S_{NB}}{\partial x^{j}},
\end{equation}
and that the symmetric part of the transport coefficient matrix $\ensuremath{L^{ij}}$
is positive semidefinite. Suppose further that the reversible part
preserves $S_{\mathrm{NB}}$, namely
\begin{equation}
\partial_{t}^{(H)}S_{NB}=0.\label{eq:SdelTimeH}
\end{equation}
Then the time evolution of $S_{\mathrm{NB}}$ is given by
\begin{equation}
\dot{S}_{NB}=\partial_{t}^{(S)}S_{NB}=L^{ij}\frac{\partial S_{NB}}{\partial x^{i}}\frac{\partial S_{NB}}{\partial x^{j}}\geq0.
\end{equation}
Thus, under these conditions, $S_{\mathrm{NB}}$ functions as a Lyapunov
function.
\end{prop}

A sufficient condition for the preservation condition \eqref{eq:SdelTimeH}
is that $S_{\mathrm{NB}}$ can be written as a function of the conserved
quantities of the reversible sector,

\begin{equation}
H_{1},H_{2},\dots,H_{N-1}.
\end{equation}
Indeed, if

\begin{equation}
S_{NB}=f(H_{1},\dots,H_{N-1})
\end{equation}
then
\begin{equation}
\partial_{t}^{(H)}S_{NB}=\sum_{m}\frac{\partial f}{\partial H_{m}}\partial_{t}^{(H)}H_{m}=0,
\end{equation}
and the preservation condition follows. In the present piston model,
\begin{equation}
S_{NB}=H_{2}-H_{1}/T_{b}
\end{equation}
so this sufficient condition is satisfied.

\LyXZeroWidthSpace{}

Near equilibrium, this condition reduces to the usual extremum condition
for a free-energy-type thermodynamic potential under the imposed bath
conditions.
\\

The thermodynamic meaning of this result can be made explicit by considering the entropy of the heat bath. Let \(E_b\) and \(S_b\) denote its energy and entropy, respectively. Since $\dot E_b=-\dot H_1$ and the bath temperature $T_b$ is constant, we have $\dot S_b=\dot E_b/T_b$. Therefore,
\begin{equation}\label{total}
\dot S_{\mathrm{tot}}=\dot S+\dot S_b
=\dot S-\frac{\dot H_1}{T_b}
=\dot S_{\mathrm{NB}}\geq 0.
\end{equation}
Thus, in the present model, the increase of the NNET dissipation potential coincides with the total entropy increase of the subsystem and the ideal heat bath. This correspondence provides a thermodynamic interpretation of $S_{\mathrm{NB}}$, rather than a new entropy-increase law.

\subsection{Linearization of the Irreversible System and Crossover of Relaxation
Modes}

In Sec. 3.3, we showed that, in the adiabatic reversible limit, the
motion near equilibrium appears as a purely oscillatory mode.

In this subsection, we examine by linearization how this oscillatory
mode is damped in the irreversible system with a heat bath and friction,
and how it competes with the thermal relaxation mode.

We denote the equilibrium point by
\begin{equation}
(V_{*},p_{*},S_{*})=(V_{*},0,S_{*}),
\end{equation}
with the equilibrium conditions
\begin{equation}
P(S_{*},V_{*})=P_{b},\qquad T(S_{*},V_{*})=T_{b}.
\end{equation}
We write small deviations from equilibrium as

\begin{equation}
V=V_{*}+\delta V,\qquad p=\delta p,\qquad S=S_{*}+\delta S.
\end{equation}
The full time evolution is

\begin{equation}
\dot{V}=\frac{A}{m}p,
\end{equation}
\begin{equation}
\dot{p}=A(P-P_{b})-\lambda p,
\end{equation}
and
\begin{equation}
\dot{S}=\kappa\frac{T_{b}-T}{T}+\frac{\lambda p^{2}}{mT}.
\end{equation}

Here, the $p^{2}$ term is second order in the deviation from equilibrium
and therefore does not contribute to the linearized dynamics.

Thus, the linearized fluctuation equation is
\begin{equation}
\frac{d}{dt}\begin{pmatrix}\delta V\\
\delta p\\
\delta S
\end{pmatrix}=M(\lambda)\begin{pmatrix}\delta V\\
\delta p\\
\delta S
\end{pmatrix},
\end{equation}
where
\begin{equation}
M(\lambda)=\begin{pmatrix}0 & A/m & 0\\
AP_{V}^{*} & -\lambda & AP_{S}^{*}\\
-\kappa T_{V}^{*}/T_{b} & 0 & -\kappa T_{S}^{*}/T_{b}
\end{pmatrix},
\end{equation}
with
\begin{equation}
P_{V}^{*}=\left.\frac{\partial P}{\partial V}\right|_{*},\quad P_{S}^{*}=\left.\frac{\partial P}{\partial S}\right|_{*},\quad T_{V}^{*}=\left.\frac{\partial T}{\partial V}\right|_{*},\quad T_{S}^{*}=\left.\frac{\partial T}{\partial S}\right|_{*}.
\end{equation}

Let $\mu_{j}(\lambda)$ denote the eigenvalues of this matrix. We
define the corresponding decay rates by
\begin{equation}
\Gamma_{j}(\lambda)=-\mathrm{Re}\,\mu_{j}(\lambda).
\end{equation}

In the reversible limit without the heat bath,$\lambda=\kappa=0$,
the eigenvalues are purely imaginary, as shown in Sec. 3.3:
\begin{equation}
\mu_{\pm}=\pm i\omega_{*}.
\end{equation}

Thus, the time evolution near equilibrium is an undamped oscillation.
For $\lambda>0$, by contrast, the eigenvalues associated with this
oscillatory mode acquire negative real parts and the motion becomes
a damped oscillation.

Under weak thermal coupling, the mechanical oscillatory mode can be
approximated by temporarily neglecting its coupling to the heat bath
and fixing $\delta S$. Then
\begin{equation}
\frac{d}{dt}\begin{pmatrix}\delta V\\
\delta p
\end{pmatrix}=M_{osc}(\lambda)\begin{pmatrix}\delta V\\
\delta p
\end{pmatrix},
\end{equation}
where
\begin{equation}
M_{osc}(\lambda)=\begin{pmatrix}0 & A/m\\
AP_{V}^{*} & -\lambda
\end{pmatrix}.
\end{equation}
This gives
\begin{equation}
\frac{d^{2}}{dt^{2}}\delta V+\lambda\frac{d}{dt}\delta V+\omega_{*}^{2}\delta V=0,\ \omega_{*}^{2}=-\frac{A^{2}}{m}P_{V}^{*}.
\end{equation}
The corresponding eigenvalues are

\begin{equation}
\mu_{\pm}=-\frac{\lambda}{2}\pm i\sqrt{\omega_{*}^{2}-\frac{\lambda^{2}}{4}}.
\end{equation}
Therefore, the decay rate of the oscillatory mode increases approximately
as

\begin{equation}
\Gamma_{{\rm osc}}=-\mathrm{Re}\mu_{\pm}\simeq\frac{\lambda}{2}.
\end{equation}

The estimate $\Gamma_{\mathrm{osc}}\simeq\lambda/2$ is a weak-coupling
and underdamped approximation obtained by neglecting the coupling
to the thermal variable. The actual decay rates and the crossover
point are determined by the eigenvalues of the full linearized matrix
$M(\lambda)$.

In the irreversible system with a heat bath, however, there exists
a nonoscillatory thermal relaxation mode in addition to the oscillatory
mechanical mode. This thermal mode describes the relaxation of the
temperature toward the bath temperature. Its relaxation rate is mainly
determined by the heat-exchange coefficient $\kappa$, and depends
only weakly on $\lambda$.

For an ideal gas, we have
\begin{equation}
S=C_{V}\ln T+R\ln V.
\end{equation}
At the equilibrium point, this gives
\begin{equation}
T_{S}^{*}=\frac{T_{b}}{C_{V}},\qquad T_{V}^{*}=-\frac{RT_{b}}{C_{V}V_{*}},\label{eq:Tstar}
\end{equation}
and
\begin{equation}
P_{S}^{*}=\frac{P_{b}}{C_{V}},\qquad P_{V}^{*}=-\frac{P_{b}}{V_{*}}\left(1+\frac{R}{C_{V}}\right).\label{eq:Pstar}
\end{equation}

In the regime where the mechanical pressure equilibration is faster
than the heat exchange with the bath, the slow thermal mode may be
estimated by imposing the quasi-mechanical equilibrium condition

\begin{equation}
\delta P\simeq0.
\end{equation}

This approximation should not be interpreted as the asymptotic limit
$\lambda\to\infty$. For extremely large friction, the piston displacement
itself may become slow because the motion is overdamped. The approximation
used here is instead intended for the intermediate-friction regime
in which the pressure imbalance relaxes on a shorter time scale than
the thermal exchange with the bath. 

Since
\begin{equation}
\delta P=P_{V}^{*}\delta V+P_{S}^{*}\delta S=0,
\end{equation}
we obtain
\begin{equation}
\delta V=-\frac{P_{S}^{*}}{P_{V}^{*}}\delta S.
\end{equation}
On the other hand, the temperature fluctuation is
\begin{equation}
\delta T=T_{V}^{*}\delta V+T_{S}^{*}\delta S.
\end{equation}
Thus,
\begin{equation}
\delta T=\left(T_{S}^{*}-T_{V}^{*}\frac{P_{S}^{*}}{P_{V}^{*}}\right)\delta S.
\end{equation}
Using the ideal-gas expressions \eqref{eq:Tstar}\eqref{eq:Pstar},
we find
\begin{equation}
\frac{P_{S}^{*}}{P_{V}^{*}}=-\frac{V_{*}}{C_{V}+R}.
\end{equation}
Therefore
\begin{equation}
\delta T=\left[\frac{T_{b}}{C_{V}}-\left(-\frac{RT_{b}}{C_{V}V_{*}}\right)\left(-\frac{V_{*}}{C_{V}+R}\right)\right]\delta S=\frac{T_{b}}{C_{V}+R}\delta S.
\end{equation}

The linearized entropy equation is
\begin{equation}
\frac{d}{dt}\delta S=-\frac{\kappa}{T_{b}}\delta T.
\end{equation}
Substituting the relation above gives
\begin{equation}
\frac{d}{dt}\delta S=-\frac{\kappa}{C_{V}+R}\delta S.
\end{equation}

Hence, the thermal relaxation rate is approximately
\begin{equation}
\Gamma_{{\rm th}}\simeq\frac{\kappa}{C_{V}+R}.
\end{equation}
Equivalently, this is the relaxation rate determined by the constant-pressure
heat capacity
\begin{equation}
C_{P}=C_{V}+R.
\end{equation}

Thus, even when $\lambda$ is increased, the relaxation rate of this
slow thermal mode changes only weakly as long as the heat-exchange
coefficient $\kappa$ is fixed.

\LyXZeroWidthSpace{}

The irreversible piston system therefore has two relevant relaxation
modes. For small $\lambda$, friction is weak, and the damping of
the mechanical oscillatory mode limits the overall relaxation. As
$\lambda$ increases within the underdamped-to-intermediate regime
considered here, the mechanical damping rate increases and can become
faster than the thermal relaxation rate. In that case, the remaining
slow relaxation is limited mainly by heat exchange with the bath.
Thus, in the parameter range examined below, the relaxation process
exhibits a crossover,
\begin{equation}
\text{friction-limited}\quad\longrightarrow\quad\text{heat-exchange-limited}.
\end{equation}

This crossover refers to a change of the rate-limiting relaxation
mechanism, not to the large-$\lambda$ asymptotic behavior of an overdamped
mechanical oscillator.

For the numerical parameters used below, namely $m=1.0,C_{V}=1.5,R=1.0,P_{b}=1.0,T_{b}=1.0,A=1$,
and $\kappa=0.20$, the thermal relaxation rate is estimated as $\Gamma_{\mathrm{th}}\simeq\kappa/(CV+R)=0.08$.
The crossover therefore occurs when the decay rate of the mechanical
oscillatory mode becomes comparable to this thermal relaxation scale.
This crossover should be distinguished from the critical damping of
a purely mechanical oscillator; it refers instead to the exchange
of the slowest relaxation mode between mechanical damping and thermal
relaxation.

\subsection{Connection to Near-Equilibrium Linear Response}

Within the linearized full dynamics, the irreversible sector has an
Onsager-type linear-response structure.

The linearized matrix $M(\lambda)$ around the equilibrium point contains
both the reversible oscillatory component originating from the Nambu
bracket and the irreversible relaxation component generated by the
dissipation potential $S_{NB}$\LyXZeroWidthSpace . Therefore, the
full matrix $M(\lambda)$ should not be identified with an Onsager
matrix. The correspondence with Onsager-type linear nonequilibrium
thermodynamics appears only in the irreversible sector of the full
linearized dynamics:
\begin{equation}
\frac{d}{dt}\left.\delta x\right|_{irr}=L_{*}\left.\nabla^{2}S_{NB}\right|_{*}\delta x.
\end{equation}
Here, $L_{*}$ is the positive-semidefinite transport coefficient
matrix evaluated at the equilibrium point, and the linearization of
$\nabla S_{NB}$ corresponds to the thermodynamic force. By contrast,
the reversible oscillatory mode does not arise from the symmetric
dissipative part of an Onsager matrix, but from the reversible sector
generated by the Nambu bracket.

Although $L^{ij}$ generally depends on the state variables, its derivative
terms do not contribute to the linearization at equilibrium because 

\begin{equation}
\left.\nabla S_{NB}\right|_{*}=0
\end{equation}

Thus, the irreversible linearized dynamics takes the form given above.

\section{Numerical Results}

We numerically examine the dynamics of the piston system introduced
above.

\begin{figure}[th]
\centering
\centering \includegraphics[scale=0.5]{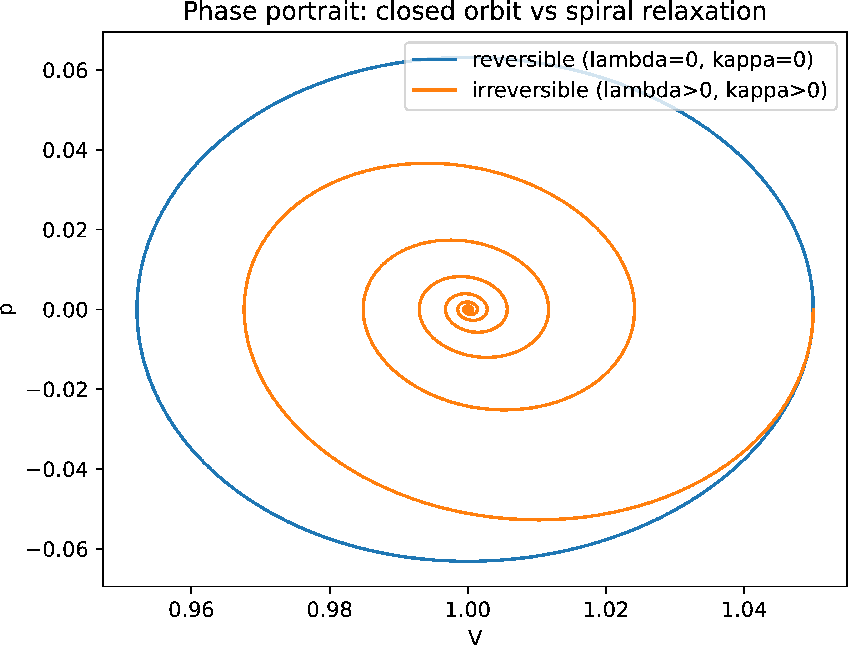}
\caption{Phase portrait of the momentum $p$ and volume $V$ for the reversible
and irreversible systems.}
\label{fig:compare_phase_portrait}
\end{figure}

Fig.~\ref{fig:compare_phase_portrait} shows the phase portrait in
the $(V,p)$ plane, with the reversible and irreversible systems superposed.
In the reversible system, the motion traces closed orbits, whereas
in the irreversible system the trajectory spirals inward because of
damping.

\begin{figure}[htbp]
\centering \includegraphics[width=0.8\textwidth,totalheight=0.5\textheight,keepaspectratio]{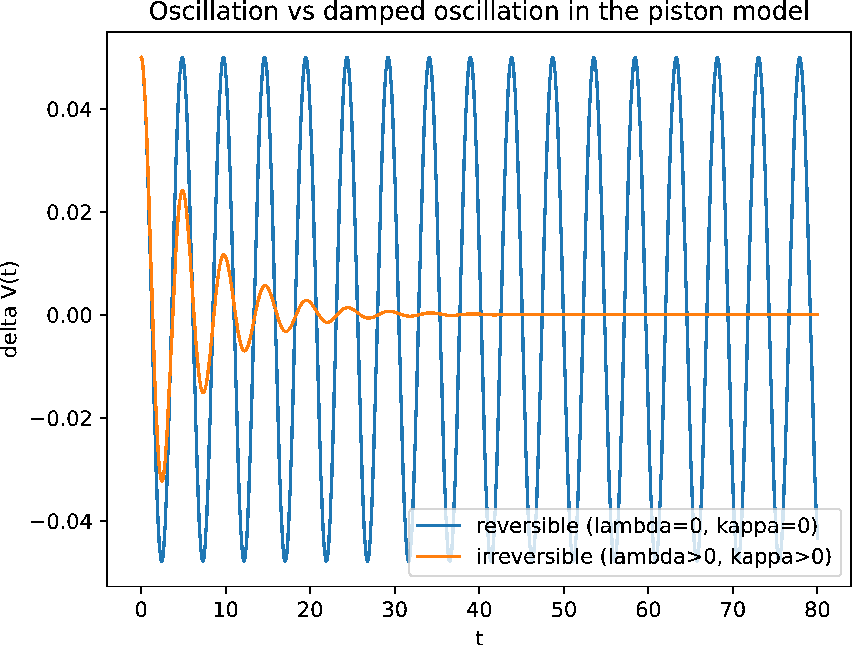}
\caption{Time evolution of the volume around the equilibrium point. In the
numerical calculation, we used $m=1.0,C_{V}=1.5,R=1.0,P_{b}=1.0,T_{b}=1.0,\lambda=0.25,\kappa=0.20$
and $A=1$, with the initial condition $(V(0),p(0),S(0))=(1.05V\ast,0,S\ast)$
. This figure corresponds to the underdamped regime, where persistent
oscillations appear in the reversible system and damped oscillations
appear in the irreversible system. In the reversible system, described
only by the Nambu bracket, the amplitude does not decay. By contrast,
when irreversible effects due to the heat bath and friction are introduced,
the system exhibits damped oscillations toward the equilibrium point.}
\label{fig:volume_oscillation} 
\end{figure}

Fig.~\ref{fig:volume_oscillation} shows the time evolution of the
volume around the equilibrium point. With only the reversible part,
the Nambu bracket generates persistent oscillations. When the irreversible
part is added, these become naturally damped oscillations.

\begin{figure}[htbp]
\centering \includegraphics[width=0.45\textwidth,totalheight=0.5\textheight,keepaspectratio]{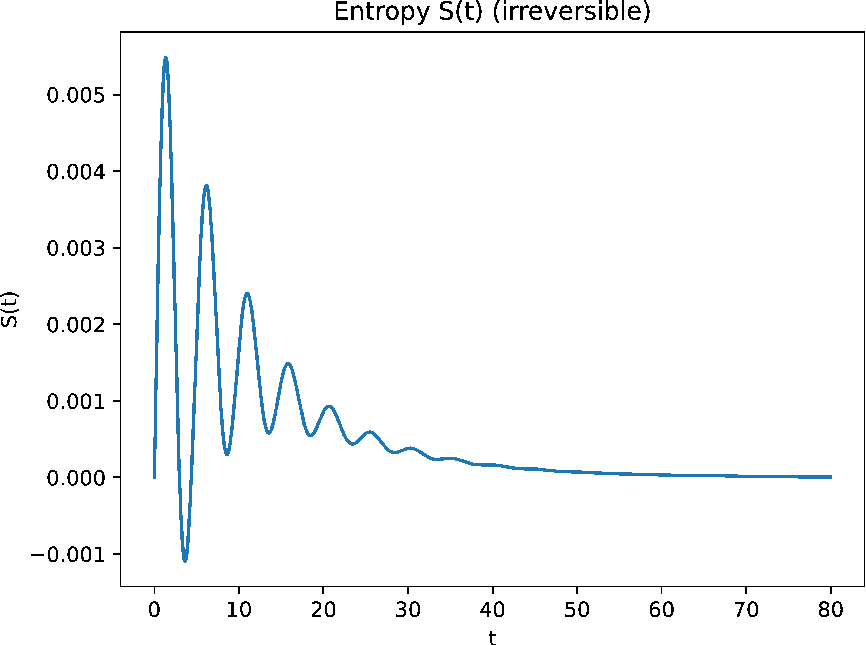}
\hfill{}\includegraphics[width=0.45\textwidth,totalheight=0.5\textheight,keepaspectratio]{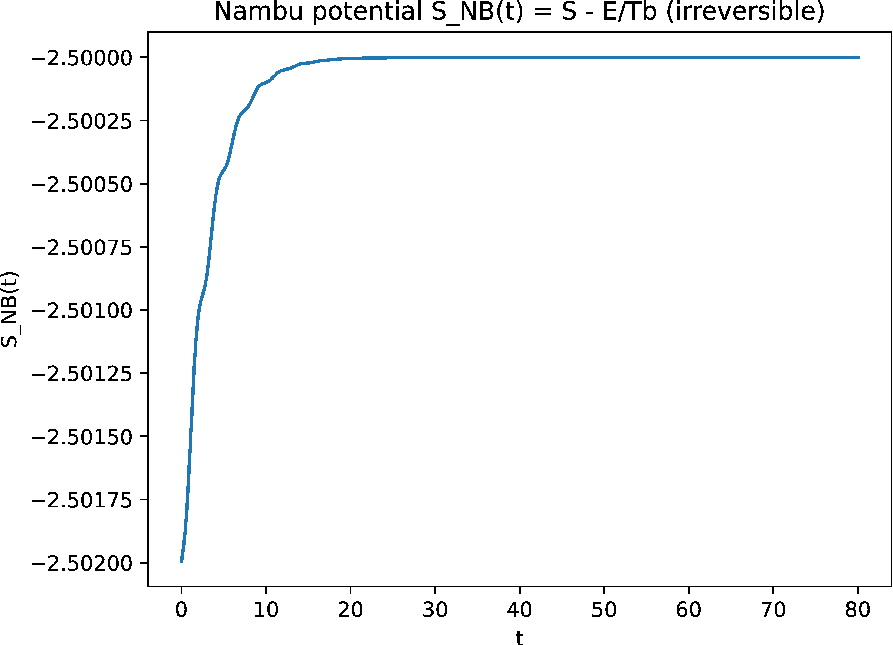}
\caption{Comparison of the time evolution of the thermodynamic entropy $S$
and the dissipation potential $S_{NB}$. In the numerical calculation,
we used $m=1.0,C_{V}=1.5,R=1.0,P_{b}=1.0,T_{b}=1.0,\lambda=0.25$
and $\kappa=0.20$. (a) The thermodynamic entropy $S$ of the system
oscillates through repeated decreases and increases due to exchange
with the heat bath, and eventually approaches its equilibrium value.
(b) By contrast, in the explicit dissipative model adopted in this
paper, the dissipation potential $S_{NB}$ receives no contribution
from the reversible part and follows a positive-semidefinite gradient
flow, so that it increases monotonically until the system reaches
equilibrium. This result does not imply that $S_{NB}$ is always monotonic
in general; rather, it shows that, under the conditions of Proposition
$1$, $S_{NB}$ functions as a Lyapunov function characterizing relaxation
toward equilibrium. As shown in Eq. (\ref{total}), this monotonic increase also corresponds to the total entropy production of the subsystem and the ideal heat bath, thereby confirming the thermodynamic interpretation of the dissipation potential.}
\label{fig:entropy_dynamics} 
\end{figure}

Fig.~\ref{fig:entropy_dynamics} provides the central numerical demonstration
of the distinction addressed in this paper by comparing the time evolution
of the thermodynamic entropy $S$ and the dissipation potential $S_{NB}$.
Since the system entropy $S$ belongs to an open subsystem, it is
allowed to oscillate, including temporary decreases. By contrast,
$S_{NB}$, constructed so as to incorporate the mechanical energy
of the system, increases monotonically. This supports its validity
as an indicator of nonequilibrium evolution within the present framework.

\begin{figure}[htbp]
\centering \includegraphics[width=0.45\textwidth,totalheight=0.3\textheight,keepaspectratio]{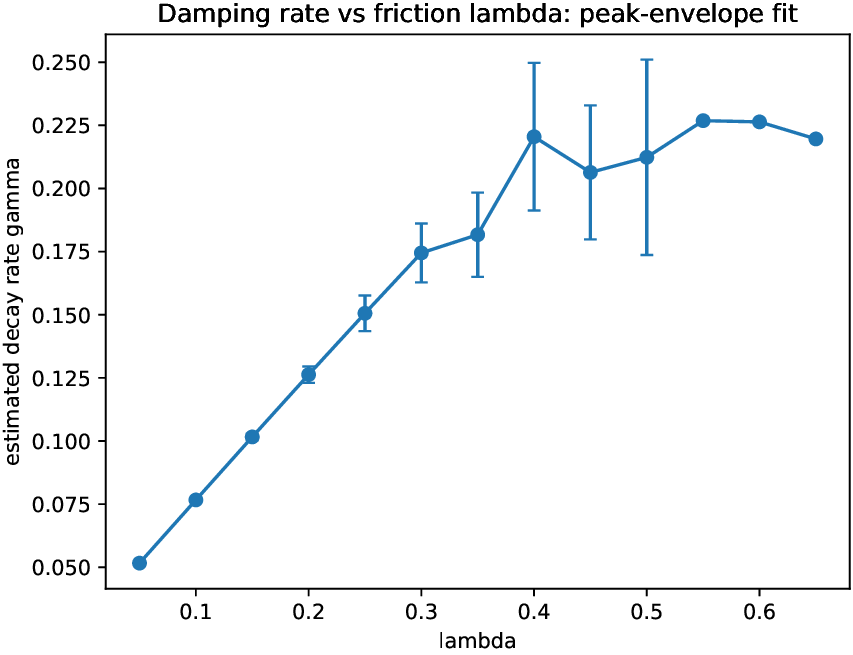}
\hfill{}\includegraphics[width=0.45\textwidth,totalheight=0.3\textheight,keepaspectratio]{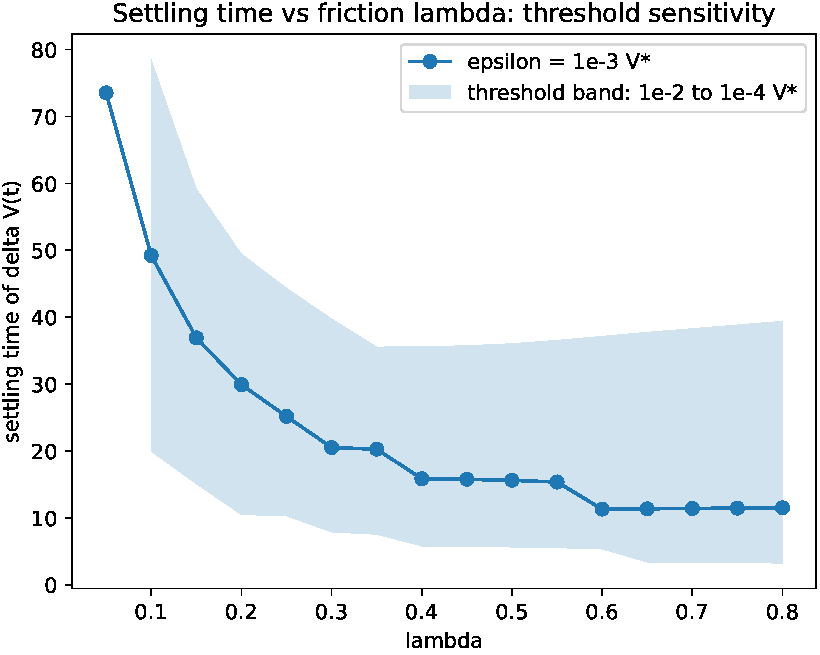}
\caption{Dependence of the damping rate $\gamma$ and the relaxation time on
the friction coefficient $\lambda$. In the numerical calculation,
we used $m=1.0,C_{V}=1.5,R=1.0,P_{b}=1.0,T_{b}=1.0$, and $\kappa=0.20$
. In the low-$\lambda$ region, $\gamma$ increases almost in proportion
to $\lambda$, reflecting the frictional damping of the mechanical
oscillation. In the intermediate-to-large-$\lambda$ region, however,
the peak-envelope estimate becomes less robust because the number
of clearly identifiable oscillation peaks decreases and the slow thermal
relaxation mode becomes relevant. Therefore, $\gamma$ should be interpreted
as a trajectory-based damping indicator rather than as a direct eigenmode
decay rate. The settling time decreases sharply for small $\lambda$,
while its decrease becomes weaker at larger $\lambda$. This saturation-like
tendency remains visible even when a threshold band is introduced.
These trajectory-based measures are consistent with the crossover
from friction-limited mechanical relaxation to heat-exchange-limited
thermal relaxation, which is examined more directly by the linearized
eigenvalue analysis in Fig.~\ref{fig:crossover}.}
\label{fig:parameter_dependence}
\end{figure}

Fig.~\ref{fig:parameter_dependence} shows the $\lambda$ dependence
of the damping rate $\gamma$ and the relaxation time. The damping
rate is estimated from the exponential envelope of the peak amplitudes
of
\begin{equation}
\delta V=V-V_{*},
\end{equation}
where $V_{*}$ is the equilibrium volume, namely
\begin{equation}
\gamma=-\frac{d}{dt}\ln|\delta V_{\mathrm{peak}}|,\ |\delta V_{\mathrm{peak}}|\sim e^{-\gamma t}.
\end{equation}
The damping rate $\gamma$ increases as $\lambda$ increases.

By contrast, the relaxation time $t$ is defined as the first time
such that 
\begin{equation}
|\delta V|<\epsilon
\end{equation}
holds thereafter, with

\begin{equation}
\epsilon=10^{-3}V_{*}.
\end{equation}

The numerical results confirm that the relaxation time decreases as
$\lambda$ increases. It is noteworthy, however, that there exists
a region in which the damping rate becomes nearly flat and shows only
weak dependence on $\lambda$.

The relatively large uncertainty observed in the intermediate range
of $\lambda$ should not be interpreted as a transition to an overdamped
regime. In the present model the friction force is linear in the momentum,
and no static-friction or stick-slip mechanism is included. Rather,
this uncertainty reflects the limited robustness of trajectory-based
estimators such as the peak-envelope fit and the settling-time criterion.
As $\lambda$ increases, the number of clearly identifiable oscillation
peaks decreases, and the long-time behavior becomes increasingly affected
by the slow thermal relaxation mode. Therefore, the damping rate estimated
from the peak envelope of $\delta V$ should be regarded as a trajectory-based
indicator, not as a direct eigenmode decay rate in the large-$\lambda$
regime.

The crossover of relaxation mechanisms is more directly characterized
by the eigenvalues of the linearized matrix. The decay rate of the
mechanical oscillatory mode increases with $\lambda$, whereas the
thermal relaxation rate is mainly controlled by the heat-exchange
coefficient $\kappa$ and remains nearly constant. Consequently, the
slowest relaxation mode changes from the friction-limited mechanical
mode at small $\lambda$ to the heat-exchange-limited thermal mode
at larger $\lambda$. This eigenvalue-based analysis provides the
clearest interpretation of the saturation-like tendency observed in
the settling-time data.

\begin{figure}
\includegraphics[scale=0.5]{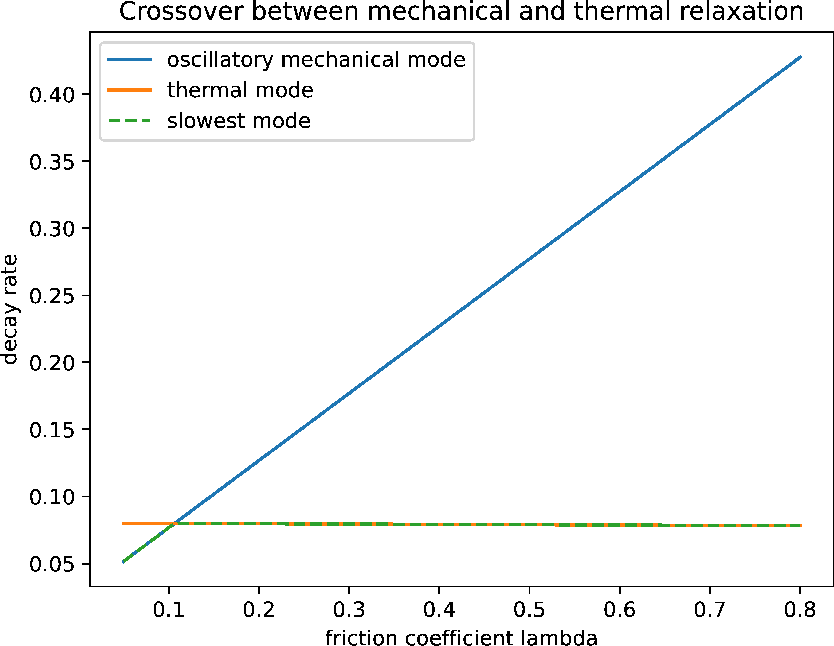}

\caption{Dependence of the decay rates obtained from the eigenvalues of the full dissipative dynamics linearized around equilibrium on the friction coefficient $\lambda$. The decay rate of the oscillatory mechanical mode increases with $\lambda$, whereas that of the thermal relaxation mode is mainly controlled by the heat-exchange coefficient $\kappa$ and depends only weakly on $\lambda$. Consequently, the slowest relaxation mode crosses over from friction-limited mechanical relaxation to heat-exchange-limited thermal relaxation. This crossover represents a change in the rate-limiting eigenmode and should not be confused with the critical damping of a purely mechanical oscillator.}\label{fig:crossover}
\end{figure}

This behavior reflects the crossover of relaxation modes discussed
in Sec. 4.3. As shown in Fig.~\ref{fig:crossover}, the $\lambda$
dependence of the decay rates obtained from the linearized irreversible
dynamics around equilibrium is nontrivial, and can be understood as
the result of competition between two relaxation modes: mechanical
damping and thermal relaxation. The decay rate of the oscillatory
mechanical mode increases with $\lambda$, whereas the decay rate
of the thermal relaxation mode is determined mainly by the heat-exchange
coefficient $\kappa$ and depends only weakly on $\lambda$. Consequently,
the slowest relaxation mode crosses over from friction-limited mechanical
relaxation to heat-exchange-limited thermal relaxation.

\section{Discussion and Conclusion}

\subsection{Conclusions of This Study}

In this paper, we investigated the thermodynamic meaning of the dissipation potential in Nambu nonequilibrium thermodynamics (NNET) using an open-piston system coupled to a pressure reservoir and a heat bath. The reversible temporal order,
namely oscillatory motion, is supported by the geometric structure
of the Nambu bracket, whereas irreversible dissipation and relaxation
are described as a gradient flow generated by the dissipation potential.
This structural decomposition into rotational flow and gradient flow
provides a natural way to describe far-from-equilibrium dynamics that
is not easily captured by frameworks based primarily on gradient-flow
descriptions, such as the maximum entropy production principle.

A central result of this study is the explicit identification of the dissipation potential as a bath-relative Massieu-type thermodynamic potential. In the present model, its time derivative coincides with the total entropy production rate of the subsystem and the ideal heat bath.
In the open piston model, the thermodynamic entropy
$S$ of the system alone can exhibit nonmonotonic time evolution through
exchange with the heat bath. By contrast, in the explicit dissipative
model constructed here, $S_{NB}$\LyXZeroWidthSpace\LyXZeroWidthSpace{}
is not the thermodynamic entropy itself, but a dissipation potential
that generates the irreversible part. Under the conditions stated
in Proposition 1, it therefore becomes a monotonic quantity. Thus,
$S_{NB}$\LyXZeroWidthSpace{} should not in general be identified with
the entropy $S$, nor should it be regarded as a universal Lyapunov
function. Rather, the monotonicity of $S_{NB}$\LyXZeroWidthSpace{}
is a conditional property that appears when its preservation by the
reversible part and the positive-semidefinite gradient-flow structure
of the irreversible part are simultaneously satisfied.

From the viewpoint of the broader NNET series, the piston model studied
here serves as a benchmark model for considering local reductions
and global obstructions that may appear in general far-from-equilibrium
systems. In a general nonlinear system, Nambu Hamiltonians and dissipation
potentials need not exist globally, and the dynamics may have to be
described by patching together local charts. In contrast, in the present
model, $H_{1}$, $H_{2}$\LyXZeroWidthSpace , and $S_{NB}$ are defined
globally, and the conditions of Proposition 1 are explicitly satisfied.
The model therefore provides a solvable reference point in which $S_{NB}$\LyXZeroWidthSpace{}
functions as a global Lyapunov function.

\subsection{Discussion: Idealization of the Pressure Reservoir}

In this paper, we idealized the pressure reservoir as a work reservoir
that supplies a constant external pressure, and assumed that the pressure
reservoir itself carries no entropy exchange,
\begin{equation}
\dot{S}_{P}=0.
\end{equation}
Under this assumption, the conserved quantity of the reversible part
closes as $H_{1}$, and the second law can be traced unambiguously
as the increase of the total entropy including the heat bath.

A more detailed modeling would be possible in which the pressure reservoir
has a finite temperature and the work received by the reservoir is
dissipated internally. In that case, however, the definition of the
reservoir entropy may depend on how far one includes the thermalization
of work inside the reservoir. For the purpose of the present paper,
namely to visualize the distinct roles of the Nambu bracket and the
dissipation potential, the idealization of the pressure reservoir
as a work reservoir is minimal and transparent.

\subsection{Application to General Open Systems Such as Chemical Reaction Networks}

The framework presented in this paper can also be applied to chemical
reaction networks that generate dissipative structures. For a constant-volume
reaction system with state variables
\begin{equation}
x=(c^{1},\dots,c^{M},S),
\end{equation}
conserved quantities such as elemental conservation laws and the total
internal energy can be assigned to the Nambu Hamiltonians $H_{i}$
. Under suitable conditions, such conserved quantities may be used to construct a Nambu representation of the rotational component of the dynamics.

For the irreversible part, let $\nu_{r}^{\alpha}$ be the stoichiometric
coefficient of reaction $r$, and let $\mu_{\alpha}$ be the chemical
potential of species $\alpha$. The reaction affinity is then defined
by

\begin{equation}
A_{r}=-\sum_{\alpha}\nu_{r}^{\alpha}\mu_{\alpha}.
\end{equation}
 Near equilibrium, the reaction flux can be written in the Onsager
form
\begin{equation}
J_{r}=L_{rs}A_{s}.
\end{equation}
When the system is coupled to chemical reservoirs, or chemostats,
one may introduce a Massieu-type generating potential that is analogous
to the piston-system expression $S_{NB}=S-H_{1}/T_{b}$ . Including
the chemical potentials of the reservoirs, this potential takes the
form
\begin{equation}
S_{NB}=S-\frac{1}{T_{b}}(U-\sum\mu_{i}^{(b)}N_{i}),
\end{equation}
where $\mu_{i}^{(b)}$ denotes the chemical potential of the reservoir.
In NNET, the irreversible part is then represented as a gradient flow
generated by this $S_{NB}$.

Complex temporal order observed in nonlinear reaction systems, such
as limit cycles in the Belousov-{}-Zhabotinsky reaction, cannot be
explained by simple gradient descent alone. From the viewpoint of
NNET, the main component responsible for temporal order is the reversible
part described by the Nambu bracket, while the irreversible part determines
its amplitude, stability, and transitions. This geometric division
of roles is the same as the structure demonstrated in the piston system
studied in this paper.

However, in such strongly nonlinear reaction systems, the global conditions
of Proposition 1 need not hold. In particular, a single $S_{NB}$
need not function as a global Lyapunov function for the entire time
evolution. Rather, through the coupling between reversible circulation
and nonlinear dissipation, $S_{NB}$ itself may exhibit nonmonotonic
time evolution. The open piston is the simplest macroscopic thermodynamic
system in which this separation can be tested explicitly.

\subsection{Perspective on Connections to Modern Nonequilibrium Statistical Mechanics}

The following discussion is not required for the macroscopic thermodynamic
formulation developed above, but indicates possible connections to
modern stochastic thermodynamics.

In recent years, stochastic thermodynamics has actively investigated
the relation between entropy production and nonconservative probability
currents in Langevin systems and Markov jump processes. The monotonic
increase of the macroscopic dissipation potential $S_{\mathrm{NB}}$
demonstrated in the present model can be understood as a structure
corresponding to the monotonic decrease of relative entropy expressed
by the Kullback-{}-Leibler divergence, or to the positivity of nonadiabatic
entropy production, in stochastic thermodynamics. Extensions toward
fluctuating Nambu nonequilibrium thermodynamics, which incorporates
microscopic fluctuations, and the role of the Nambu bracket in nonequilibrium
steady states (NESSs) between multiple heat baths have been discussed
in Ref.\cite{Katagiri_2022_Fluctuating}. The macroscopic model studied
in the present paper can be regarded as a deterministic limit of such
a formulation. A systematic clarification of the relation between
these descriptions remains an important subject for future work.

When the theory is extended to a stochastic description, $\ensuremath{S_{\mathrm{NB}}}$may
also be related to Shannon entropy and self-information. For example,
if a Nambu-type distribution of the form
\begin{equation}
\phi\propto\exp\left[\frac{S_{\mathrm{NB}}}{\alpha k_{B}}\right]
\end{equation}
appears as a stationary distribution of Fokker-{}-Planck type, then
the self-information $-\log\phi$ corresponds to $-\ensuremath{S_{\mathrm{NB}}/(\alpha k_{B})}$,
up to an additive normalization constant. This relation is an important
issue in extending the deterministic macroscopic model treated in
this paper to stochastic NNET.

\subsection{Future work}

Another natural extension is to replace the ideal-gas equation of
state by an interacting-fluid equation of state, such as a van der
Waals fluid. The Nambu formulation of the reversible sector and the
bath-relative potential $S_{NB}=S-H_{1}/T_{b}$ can be written formally
for a general thermodynamic potential $U(S,V)$. Therefore, the main
geometric structure is not restricted to an ideal gas. However, the
explicit linearized relaxation rates and the numerical examples presented
in this paper use the ideal-gas equation of state. For interacting
fluids, the derivatives of $P(S,V)$ and $T(S,V)$ would be modified,
and additional issues such as metastability, phase coexistence, and
possible spatial inhomogeneities near coexistence regions may have
to be considered. A systematic analysis of such interacting-fluid
extensions is left for future work.

\section*{Acknowledgments}

The author would like to thank Akio Sugamoto and Yoshiki Matsuoka
for their collaboration in the development of Nambu nonequilibrium
thermodynamics and for valuable comments on the present work. The
author is deeply grateful to Tatsuaki Wada for carefully reading the
manuscript and for many helpful discussions. The author also thanks
Shiro Komata for reading the manuscript and for constructive questions,
in particular on the relation to Onsager theory and the meaning of
the dissipation potential $S_{\mathrm{NB}}$. The author is also grateful
to the members of the Field Theory Seminar at the Open University
of Japan for stimulating discussions.

\bibliographystyle{unsrt}
\bibliography{piston}

\end{document}